\documentstyle[twocolumn,floats,aps]{revtex}
\input psfig
\begin{document}
\draft
\title{A Solvable Model for Spatiotemporal Chaos }
\author{R. O. Grigoriev and H. G. Schuster*}
\address{Condensed Matter Physics 114-36 and Neural Systems Program
139-74 \\ California Institute of Technology, Pasadena CA 91125}
\date{\today }
\maketitle

\begin{abstract}
 We show that the dynamical behavior of a coupled map lattice where the
individual maps are Bernoulli shift maps can be solved analytically for
integer couplings. We calculate the invariant density of the system and
show that it displays a nontrivial spatial behavior. We also introduce and
calculate a generalized spatiotemporal correlation function.
 \end{abstract}


\section{Introduction}

The study of temporal chaos in low-dimensional systems, some of which can
be described by low-dimensional maps \cite{frish,schuster}, was extremely
beneficial for the understanding of turbulence. In 1984 coupled map
lattices were introduced into the physical literature as a tool for
studying spatiotemporal chaos in spatially extended, i.e. high-dimensional
systems \cite{waller}. They consist of spatially coupled low-dimensional
maps and represent dynamical systems that are discrete in space and time,
but continuous in the state variable. They serve as models for coupled
Josephson junctions, excitable media, population dynamics, neural dynamics
and turbulence \cite{kaneko}. Although Bunimovich and Sinai mathematically
proved a number of statements regarding the appearance of coherent
structures from spatiotemporal chaos \cite{sinai}, most results in the
field have been obtained by numerical simulations \cite{kaneko,politi}. 
While the study of temporal chaos has greatly profited from the existence
of simple maps like the Bernoulli shift map and the cat map
\cite{schuster,arnold}, which can be solved explicitly (for integer
expansion rates), thereby making the mechanisms of mixing and temporal
chaos understandable, no investigation of this type has been provided up
to now for the problem of spatiotemporal chaos. Here we present a solvable
model for spatiotemporal chaos which can be solved in the sense of the
Bernoulli map. 

We start from the model equations \cite{waller}
 \begin{equation}
 \label{eq_cml_gen}
 z_{i}^{t+1}=(1-\varepsilon )f(z_{i}^{t})+\frac{\varepsilon }{2}\left[
 f(z_{i+1}^{t})+f(z_{i-1}^{t})\right],
 \end{equation}
 where the index $i$ runs over the $N$ sites of a discrete lattice,
$\varepsilon$ measures the strength of the spatial coupling between
neighboring sites and $f(z_i^t)$ is a local map which determines the
evolution of the continuous variable $z_i^t$ at discrete time-steps
$t=0,1,2,\cdots$. The extension of eq. (\ref{eq_cml_gen}) to more than
nearest neighbor coupling and arbitrary dimensions is straightforward. We
first write it in the new variables $x_i^t=f(z_i^t)$ as
 \begin{equation}
 \label{eq_cml}
 x_{i}^{t+1}=f\left[(1-\varepsilon)x_{i}^{t}+\frac{\varepsilon}{2}
 \left(x_{i+1}^{t}+x_{i-1}^{t}\right) \right].
 \end{equation}

In order to obtain a solvable model we choose $f(x)=(ax)\,{\rm mod}\;1$,
i.e. the Bernoulli shift map with integer stretching factor $a$. To make
our approach for the spatially extended system more transparent we will
calculate in Section \ref{sec_single} the invariant density and a temporal
correlation function for the single Bernoulli shift map. In section
\ref{sec_lattice} we will show that the coupled map lattice is solvable
for special values of the coefficients $a$ and $\varepsilon$ in eq. 
(\ref{eq_cml}) and we will also calculate the invariant density and a
spatial and a spatiotemporal correlation functions. Finally in section
\ref{sec_discuss} we discuss our results and indicate directions of
further research. 

\section{Properties of a single Bernoulli shift map}
\label{sec_single}

First we recall that the single map $x^{t+1}=(ax^{t})\,{\rm mod}\;1$
can be solved as $x^{t}=(a^{t}x^{0})\,{\rm mod}\;1$
because
\begin{eqnarray}
 x^{2}=&&\left\{ a\left[ (ax^{0})\,{\rm mod}\;1\right] \right\}
 {\rm mod}\;1\cr =&&\left\{ a\left[ ax^{0}-k^{0}\right] \right\} 
 \,{\rm mod}\;1=(a^{2}x^{0})\,{\rm mod}\;1 
 \end{eqnarray}
 where $k^0$ is an integer which represents the action of the modulo
function. For the last equality sign in to hold we needed the fact that
$a$ is an integer such that $a k^0$ becomes again an integer, which can be
dropped within the last modulo function. 

Since the modulo function confines the variable $x^{t}$ to a circle we
could view the Bernoulli shift map as a linear map $x^{t+1}=ax^{t}$
where the variables live on a unit circle, i.e. on a 1-torus. We
shall see below that we can view our coupled map system as a linear map
acting on variables confined to an $N$-torus, where $N$ is the number of
lattice sites.

The invariant density $\rho(x)$ of the simple Bernoulli shift map measures
the distribution of $x$ values on the attractor generated by the map and
is well known to be a constant \cite{schuster}. We can obtain this result
by noting that $\rho (x)$ is defined on an unit circle, i.e. it is
periodic in $x$ and therefore can be represented as a Fourier series
 \begin{equation}
 \label{eq_four_sing}
 \rho (x)=\sum_k\hat{\rho}(k)e^{2\pi ikx},
 \end{equation}
 where $k$ takes only integer values $k=0,\pm 1,\pm 2,\cdots$. The
invariant density $\rho(x)$ evolves from an initial distribution
$\rho^0(x)$ according to the Frobenius-Perron equation \cite{schuster}
 \begin{equation}
 \label{eq_dens_sing}
 \rho^t\left( x\right)=\int_{0}^{1}dx^{\prime }\delta
 \left[ x-\left( a^{t}x'\right)\,
 \mathop{\rm mod}\;1\right] \rho^0\left( x^{\prime }\right) 
 \end{equation}
 and is defined as the long-time limit $\rho(x)=\lim_{t\rightarrow \infty}
\rho^t(x)$. In order to solve eq. (\ref{eq_dens_sing}) we use
(\ref{eq_four_sing}) and the fact that the Bernoulli shift
map becomes a linear map on a torus, i.e. $\exp(2\pi i\left[(a^tx)\,{\rm
mod}\;1\right])=\exp(2\pi ia^{t}x)$, and obtain
 \begin{equation}
 \hat{\rho}^t(k)=\hat{\rho}^0(a^t k). 
 \end{equation}

If we make the reasonable assumption that the initial distribution
$\rho^0(x)$ is non-singular, then $\lim_{k\rightarrow\pm\infty}
\hat{\rho}^0(k)=0$. This means that in eq. (\ref{eq_four_sing}) all
Fourier coefficients of $\hat{\rho}^t(k)$ tend to zero in the
infinite-time limit, except the one, which belongs to $k=0$. Since
 \begin{equation}
 \hat{\rho}^0(0)=\int_0^1 dx\rho^0(x)=1
 \end{equation}
 this yields $\hat{\rho}(k)=\delta_{k,0}$ and $\rho(x)=1$.

In a similar fashion we can now define and calculate the time correlation
function on the 1-torus. The usual time correlation function is defined
via 
 \begin{equation}
 \left\langle xx^t\right\rangle=\int_0^1 dx\rho(x)xf^t(x),
 \end{equation}
 where the time evolution of $x$ is given by the map $f(x)$, but here we
introduce the time correlation function $G(t)$
 \begin{equation}
 \label{eq_sing_cor}
 G(t)=\int_0^1 dx^0\rho(x^0)e^{2\pi i(x^0-x^t)},
 \end{equation}
where 
 \begin{equation}
 \label{eq_sing_sol}
 x^{t}=(a^{t}x^0)\,{\rm mod}\;1,
 \end{equation}
which respects the fact that the variable $x$ is an angular variable on a
torus \cite{jose}.

For $\rho(x)=1$ we obtain from eqs. (\ref{eq_sing_cor},\ref{eq_sing_sol}) 
 \begin{equation}
 G(t)=\int_{0}^{1}dx\rho \left( x\right) 
 e^{2\pi i\left( 1-a^{t}\right)x}=\delta_{1,a^{t}}.
 \end{equation}

In the following section we will demonstrate what changes have to be made
in order to compute in a similar fashion as above the solution of the
dynamical equations, the invariant density and the time correlation
function for our coupled map lattice.

\section{A lattice of coupled Bernoulli shift maps.}
\label{sec_lattice}

For the Bernoulli shift map $f(x)=(ax)\,{\rm mod}\;1$ the time evolution
for the variables $x_{i}^{t}$ of the coupled map lattice becomes according
to eq. (\ref{eq_cml}) 
 \begin{equation}
 \label{eq_map_nnc}
 x_{i}^{t+1}=\left(a\left[(1-\varepsilon)x_{i}^{t}+\frac{\varepsilon}{2}
 (x_{i+1}^{t}+x_{i-1}^{t})\right]\right)\,{\rm mod}\;1.
 \end{equation}

If the parameters $a=m+2n$ and $\varepsilon=2n/(m+2n)$ are such that both
$(1-\varepsilon)a$ and $a\varepsilon/2$ take integer values $m$ and $n$,
then the equation of motion for the coupled map system can be written in
the compact form
 \begin{equation}
 \label{eq_map_mat}
 x_{i}^{t+1}=\left(\sum_{j}A_{ij}x_{j}^{t}\right){\rm mod}\;1,
 \end{equation}
where the coupling matrix $A$ has integer elements
 \begin{equation}
 \label{eq_mat_coup}
 A_{ij}=m\delta_{i,j}+n(\delta_{i,i+1}+\delta_{i,i-1}). 
 \end{equation}

We will now free ourselves from the specific form (\ref{eq_mat_coup}) for
$A_{ij}$, which was physically motivated by the nearest neighbor lattice
model (\ref{eq_map_nnc}) and show that eq. (\ref{eq_map_mat}) can be
solved for {\em any} matrix $A$, which has {\em integer valued elements}
$A_{ij}$. In order to see this we write (\ref{eq_map_mat}) in vector
notation as
 \begin{equation}
 \label{eq_map_vec}
 {\bf x}^{t+1}=\left( A{\bf x}^{t}\right){\rm mod}\;1,
 \end{equation}
 where ${\bf x}^{t}=(x_{1}^{t},\cdots,x_{N}^{t})$ and the modulo is taken
for each component of the vector $A{\bf x}^t$. Then we obtain by iterating
from the initial condition: 
 \begin{equation}
 {\bf x}^{t+1}=\left(A{\bf x}^t\right){\rm mod}\;1
 = A{\bf x}^t-{\bf k}^t,
 \end{equation}
where ${\bf k}^t$ is a vector with integer components which represents
the action of the modulo function. This yields 
 \begin{eqnarray}
 \label{eq_mod_iter}
 {\bf x}^{t+2} = &&\left( A{\bf x}^{t+1}\right) {\rm mod}\;1
 =\left( A\left[ A{\bf x}^t-{\bf k}^t\right] \right){\rm mod}\;1
 \cr = &&\left( AA{\bf x}^t-A{\bf k}^t\right){\rm mod}\;1
 =\left( A^2{\bf x}^t\right){\rm mod}\;1,
 \end{eqnarray}
 where the last equality sign only holds because all elements of the
matrix $A$ are integers, such that $A{\bf k}^t$ is a vector with integer
components which can be dropped under the last modulo function. Since
(\ref{eq_mod_iter}) holds for any $t$, we obtain the closed-form solution
as a function of the initial value
 \begin{equation}
 \label{eq_solution}
 {\bf x}^{t}=\left( A^t{\bf x}^0\right){\rm mod}\;1.
 \end{equation}

Eq. (\ref{eq_mod_iter}) shows that we can solve not only our coupled map
lattice problem (\ref{eq_map_nnc}), but all linearly coupled systems,
where the coupling occurs via a matrix $A$ with integer elements $A_{ij}$
and the nonlinearity is provided by the modulo function. The solution can
be obtained by first solving the linear problem, i.e. by obtaining
$A^t{\bf x}^0$ and then taking the modulo, which is the same as having the
linear map acting on an $N$-torus in analogy to the famous Arnold's cat
map in two dimensions \cite{arnold}.

Next we investigate the invariant density and the spatiotemporal
correlation function of the coupled map lattice. The first quantity gives
us information about the measurable time averaged spatial structures in
the system and the second one tells us about the measurable spatiotemporal
structures.

\subsection{The Invariant Density}
\label{sec_inv_dens}

The invariant density $\rho({\bf x})$ yields the distribution of points on
the attractor generated by the map ${\bf x}^{t+1}=\left(A{\bf x}^t\right)
{\rm mod}\;1$. By starting from an initial distribution $\rho^0({\bf x})$
it could be obtained as the infinite-time limit of $\rho^t({\bf x})$ in
the Frobenius-Perron equation
 \begin{equation}
 \label{eq_dens_ev}
 \rho^t({\bf x}) = \int d{\bf x}' \delta \left[{\bf x}-
 \left( A^t{\bf x}'\right){\rm mod}\;1\right] \rho^0({\bf x}')
\end{equation}

Since all quantities involved in eq. (\ref{eq_dens_ev}) are periodic on
an $N$-torus, the Fourier decomposition of $\rho^t({\bf x})$ contains
only wavevectors ${\bf k}$ with integer components, i.e.
 \begin{equation}
 \label{eq_dens_four}
 \rho^t({\bf x})=\sum_{\bf k}\hat{\rho}^t({\bf k})
 e^{2\pi i\,{\bf k}\cdot {\bf x}}.
 \end{equation}
 By using the equality $\exp(2\pi i\,{\bf k}\cdot[(A^t{\bf x})\,{\rm
mod}\;1]) = \exp(2\pi i[(A^t)^T{\bf k}]\cdot{\bf x})$ eq. 
(\ref{eq_dens_four})  yields
 \begin{equation}
 \label{eq_fou_comp}
 \hat{\rho}^t({\bf k})=
 \hat{\rho}^0\left(\left(A^t\right)^T{\bf k}\right).
 \end{equation}
 If the initial distribution $\rho^0({\bf x})$ is non-singular, all
Fourier coefficients vanish for large values of the wavevector, i.e.
 \begin{equation}
 \lim_{|{\bf k}|\rightarrow \infty }\hat{\rho}^0({\bf k})=0.
 \end{equation}

For a completely expanding map, where all eigenvalues of the matrix $A$
have an absolute value larger than one, $\lim_{t\rightarrow\infty}\left(
A^{t}\right)^{T}{\bf k}=\infty$ for each ${\bf k}\ne 0$ and the only
non-vanishing Fourier component becomes
 \begin{equation}
 \hat{\rho}^0(0) =\int d{\bf x}\rho^0({\bf x})=1, 
 \end{equation}
which yields a constant invariant density
 \begin{equation}
 \rho({\bf x})=1.
 \end{equation}
 This result is completely analogous to the single map case.  However we
may obtain different results for the invariant density if there are
contracting directions in the phase space. 

Indexing the stable and unstable eigenvalues $\lambda$ and right (left)
eigenvectors ${\bf e}$ ($\tilde{\bf e}$) of the matrix $A^T$ with indices
``s'' and ``u'' respectively, we have
 \begin{equation}
 \label{eq_eig_exp}
 \left(A^t\right)^T{\bf k}=
 \sum_s\lambda_s^t(\tilde{\bf e}^s\cdot{\bf k}){\bf e}^s+
 \sum_u\lambda_u^t(\tilde{\bf e}^u\cdot{\bf k}){\bf e}^u. 
\end{equation}
 According to the above, we will only obtain results which differ from the
trivial expanding case, if there exists at least one ${\bf k}\ne 0$, such
that its components along the unstable directions are all zero, i.e.  it
is contained in the stable manifold $W^s$ of the fixed point ${\bf k}=0$
of the ``conjugate'' map
 \begin{equation}
 \label{eq_conj_map}
 {\bf k}^{t+1}=A^T{\bf k}^t.
\end{equation}

In fact the above argument does not take into account one particular
feature of the system. Specifically, the map (\ref{eq_conj_map}) might not
be hyperbolic, i.e. it might possess the {\em central} manifold, defined
by the eigenvectors corresponding to $|\lambda|=1$. If this is the case
(and it is for the coupled map lattice (\ref{eq_map_nnc}) as we shall see
below), we assume that the invariant density, that we calculate
corresponds to the {\em physical} (or Kolmogorov) {\em measure}
\cite{eckmann}. The latter is calculated by introducing a small amount of
noise into the system and then taking the zero-noise limit. 

Assuming that the physical measure is unique, we can obtain the invariant
density starting from $\rho^0({\bf x})=1$, which is equivalent to
$\hat{\rho}^0({\bf k})=\delta_{{\bf k},0}$. Then only the stable manifold
contributes to the invariant density and the central manifold can be
treated as the unstable one. This essentially means, that the averaging
over the central directions is accomplished by the infinitely small noise
present in the system.

On the other hand, eqs. (\ref{eq_fou_comp},\ref{eq_eig_exp}) tell us that
it is not enough to have contracting eigenvalues in order to get a
non-constant invariant density.

\begin{figure} 
\centering
\mbox{
\psfig{figure=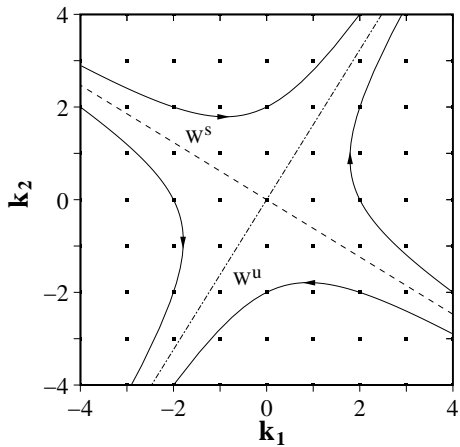,width=3in}}
\vskip 5mm
 \caption{ The cat map: none of the integer-component wavevectors lies on 
the stable manifold $W^s$ of the fixed point ${\bf k}=0$. }
 \label{fig_cat}
\end{figure}

Let us first consider the case with a single stable direction ${\bf e}^s$. 
Since all components $k_i$ of a vector ${\bf k}$ are integer, it is
contained within the stable manifold $W^s$ only if $\nu{\bf k}={\bf e}^s$.
This in turn means that the components $\{e^s_1,\cdots,e^s_N\}$ should be
mutually rational, i.e. 
 \begin{equation}
 e_1^s:e_2^s:\cdots:e_N^s=k_1:k_2:\cdots:k_N.
 \end{equation}

An example where we have one contracting and one expanding direction is
the cat map
 \begin{equation}
 A=\left(\matrix{1 & 1 \cr 1 & 2}\right).
 \end{equation}
 Although the eigenvalue corresponding to the contracting direction is
$\lambda_s=(3-\sqrt{5})/2<1$, this map still has a constant invariant
density because the components of the eigenvector ${\bf e}^s=(2,1-
\sqrt{5})$ belonging to $\lambda_s$ have a non-rational ratio (see
fig. \ref{fig_cat}) leading to
 \begin{equation}
 \lim_{t\rightarrow \infty }\hat{\rho}^t({\bf k})=\delta_{{\bf k},0}.
 \end{equation}

Generally, in order to get a non-constant invariant density our model must
possess a stable manifold, which in turn should contain at least one
vector with mutually rational components. Every vector ${\bf k}^*$ with
integer components, which is pulled in the long time limit into the
origin, according to eq. (\ref{eq_eig_exp}),
 \begin{equation}
 \label{eq_pull_in}
 \lim_{t\rightarrow \infty }\hat{\rho}^t({\bf k}^*)=
 \lim_{t\rightarrow \infty }\hat{\rho}^0
 \left(\left( A^{t}\right)^T {\bf k}^*\right)=\hat{\rho}^0(0)=1
 \end{equation}
 can be represented as a linear combination of a (usually small) number
of basis vectors ${\bf f}^j$, $j=1,\cdots,M$ with integer coefficients
$n_j$, i.e. ${\bf k}^*=\sum_j n_j{\bf f}^j$. The invariant density will
only contain non-vanishing Fourier components with such wavevectors ${\bf
k}^*$, all with weight 1:
 \begin{eqnarray}
 \label{eq_inv_dens}
 \rho({\bf x})&=&\sum_{{\bf k}^*}e^{2\pi i\,{\bf k}^*\cdot{\bf x}}
 =\prod_{j=1}^M\sum_{n_j}e^{2\pi i n_j\,{\bf f}^j\cdot{\bf x}}\cr
 &=&\prod_{j=1}^M\sum_{n_j}\delta({\bf f}^j\cdot{\bf x}-n_j).
 \end{eqnarray}

\subsection{The Coupled Map Lattice}
\label{sec_coup_map}

Up to now our conclusions have been completely general for any coupling
matrix $A$ with integer elements. Let us now consider the condition
(\ref{eq_pull_in}) in more detail for our 1-dimensional nearest neighbor
model (\ref{eq_map_nnc}). The corresponding matrix (\ref{eq_mat_coup}) can
be diagonalized by Fourier transformation in the space variables $i$,
leading for periodic boundary conditions to eigenvalues
 \begin{equation}
 \label{eq_evalue}
 \lambda_q=m+2n\cos(q)
 \end{equation}
 and the corresponding eigenvectors
 \begin{eqnarray}
 \label{eq_evector}
 {\bf e}_c^q&=&N^{-1}(\cos(q),\cos(2q),\cdots,\cos(Nq)),\cr
 {\bf e}_s^q&=&N^{-1}(\sin(q),\sin(2q),\cdots,\sin(Nq)),
 \end{eqnarray}
 where $q=2\pi p/N$ and $p=0,\cdots,N/2$.

Of these only a few have mutually rational components. For instance, both
$\cos(q):1$ and $\sin(2q):\sin(q)$ are rational only if $\cos(q)$ is
rational, which immediately restricts the allowed wavevectors $q=2\pi p/N$
to a set of 5 values: $q^*=0,\pi/3,\pi/2,2\pi/3,\pi$.  Each $q^*$
generates several basis vectors, provided $|\lambda_{q^*}|<1$: 
 \begin{eqnarray}
 \label{eq_basis}
 {\bf f}^0=&&(1,\cdots,1);\cr
 {\bf f}^{\pi/3}_1=&&(1,-1,-2,-1,1,2,\cdots,2);\cr
 {\bf f}^{\pi/3}_2=&&(-1,-2,-1,1,2,1,\cdots,1);\cr
 {\bf f}^{\pi/2}_1=&&(0,-1,0,1,\cdots,1);\cr
 {\bf f}^{\pi/2}_2=&&(1,0,-1,0,\cdots,0);\cr
 {\bf f}^{2\pi/3}_1=&&(-1,-1,2,\cdots,2);\cr
 {\bf f}^{2\pi/3}_2=&&(-1,2,-1,\cdots,-1);\cr
 {\bf f}^\pi=&&(-1,1,\cdots,1).
 \end{eqnarray}

Rationality of $\cos(q^*)$ in not an unexpected result, e.g. choosing
$\cos(q^*)=-m/2n$ results in the eigenvalue $\lambda_{q^*}=0$, according
to (\ref{eq_evalue}), which requires
 \begin{equation}
 ({\bf f}\cdot{\bf x}^t)\,{\rm mod}\;1=0,\quad \forall t>0,
 \end{equation}
 where we defined ${\bf f}=\kappa_{q^*}{\bf e}^{q^*}$ with
$\kappa_{q^*}=N$ if $q^*=0,\pi/2,\pi$ and $2N$ otherwise. This in turn,
requires $\rho({\bf x})\sim \delta(({\bf f}\cdot{\bf x})\,{\rm mod}\;1)$,
which is seen to be the case by rewriting (\ref{eq_inv_dens}) as
 \begin{equation}
 \rho({\bf x})=
 \prod_{j=1}^M\delta(({\bf f}^j\cdot{\bf x})\,{\rm mod}\;1).
 \end{equation}

It is useful to define the projection of the invariant density $\rho({\bf
x})$ on a chosen direction ${\bf g}$:
 \begin{equation}
 \label{eq_project}
 \rho_{\bf g}(s)=\int\delta(s-{\bf g}\cdot{\bf x})\rho({\bf x})\,d{\bf x}.
 \end{equation}
 For example, if $g_i=\delta_{ij}$, eq. (\ref{eq_project}) gives
the distribution of the $j$-th map variable $\rho(x_j)=1$. 

If ${\bf g}$ coincides with one of the basis directions, i.e. ${\bf
g}=\nu{\bf f}^l$ for some $l$, the projection
 \begin{eqnarray}
 \label{eq_delta}
 \rho_{\bf g}(s)&=&
 \int_{I^N}\delta(s-\nu{\bf f}^l\cdot{\bf x})
 \prod_{j=1}^M\sum_p\delta(p-{\bf f}^j\cdot{\bf x})\,d{\bf x}\cr
 &=&\sum_p D_p\,\delta(s-\nu p),
 \end{eqnarray}
 ($I^N$ denotes the unit $N$-dimensional cube) becomes singular: we get a
series of $\delta$-functions with an envelope
 \begin{equation}
 D_p=\int_{I^N}\delta(p-{\bf f}^l\cdot{\bf x})
 \prod_{j\ne l}\sum_n\delta(n-{\bf f}^j\cdot{\bf x})\,d{\bf x}.
 \end{equation}
 Otherwise, the projection (\ref{eq_project}) is a continuous,
non-singular function of parameter $s$. In other words, only the
projection on the directions defined by the basis vectors ${\bf f}^j$ is
singular.

\begin{figure*}
\centering 
\mbox{ 
\psfig{figure=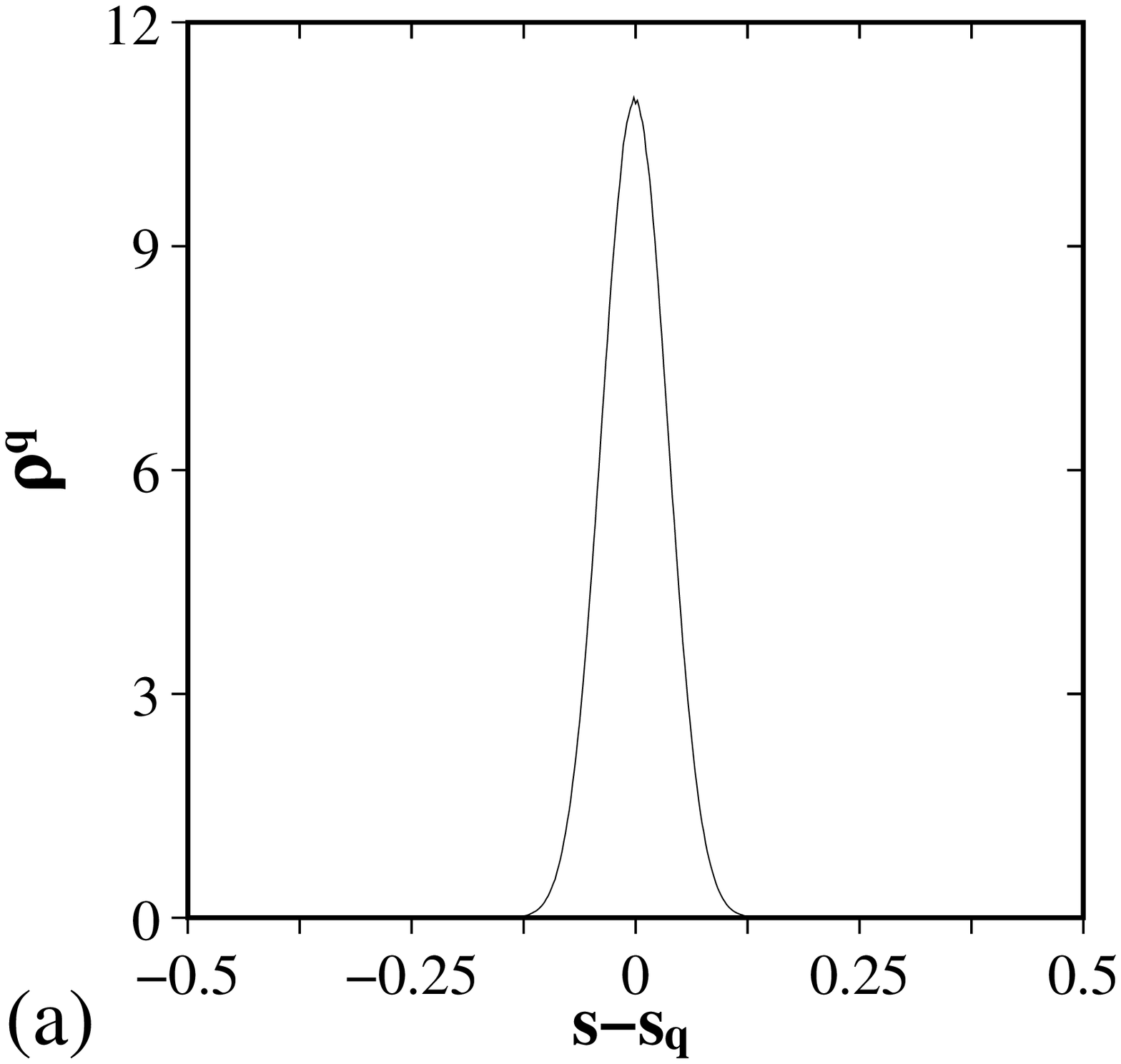,width=3in}\qquad
\psfig{figure=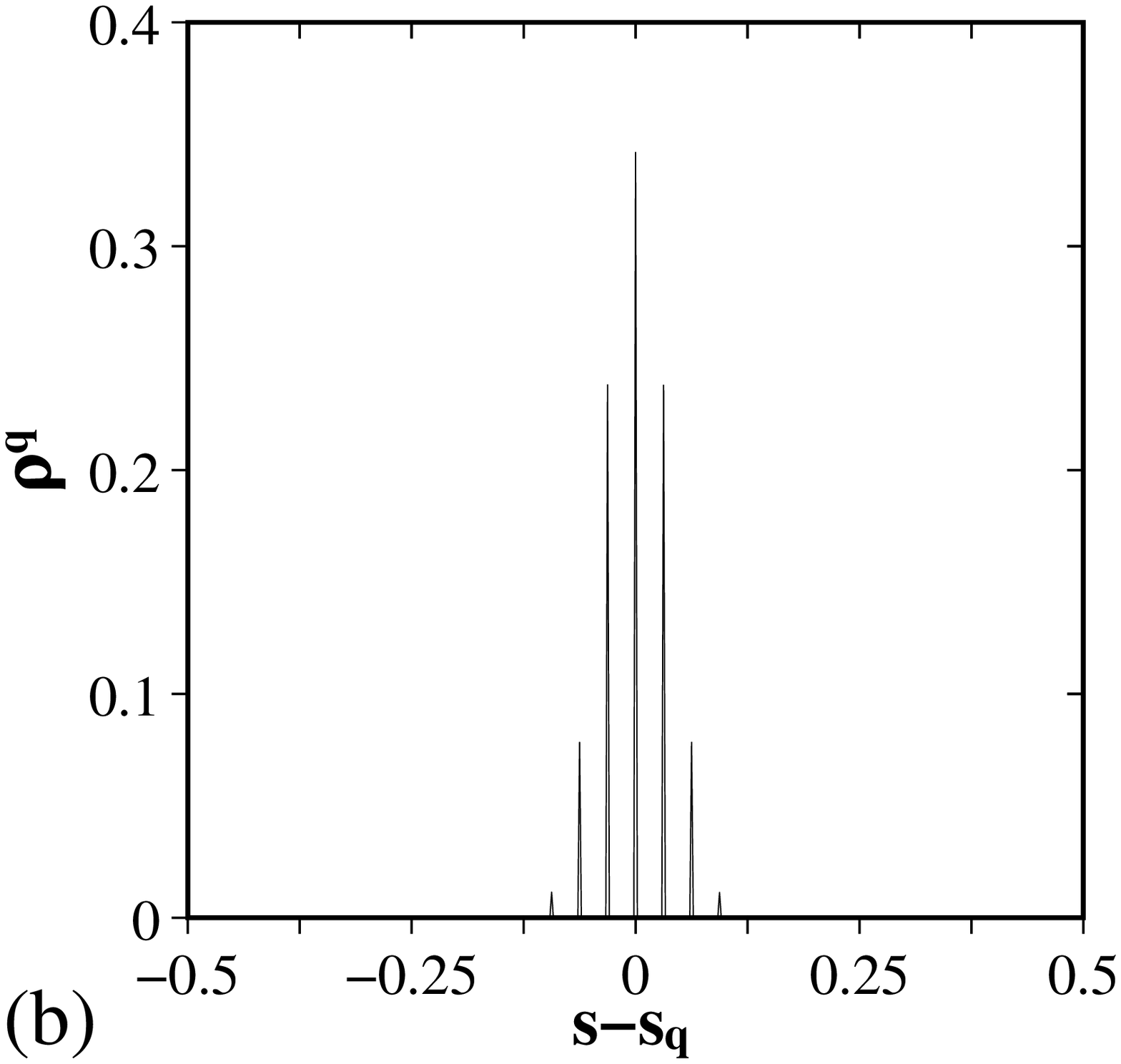,width=3in}}
\vskip 5mm
 \caption{ Projection of the invariant density $\rho^q(s-s_q)$: (a) for
$q\ne q^*$, arbitrary $\lambda_q$ and also for $q=q^*$, $|\lambda_q|>1$
and (b)  for $q=q^*$, $|\lambda_q|<1$. We used $N=32$.}
 \label{fig_narrow}
\end{figure*}

In particular, the eigenvector ${\bf e}^q$ defines a basis direction ${\bf
f}^j$ if and only if the projection (\ref{eq_project}) on this
eigenvector (we define $\rho^q(s)=\rho_{\bf g}(s)$ for ${\bf g}={\bf
e}^q$),
 \begin{equation}
 \label{eq_eig_proj}
 \rho^q(s)=
 \int_{I^N}\delta(s-{\bf e}^q\cdot{\bf x})
 \prod_{j=1}^M\sum_{p_j}\delta(p_j-{\bf f}^j\cdot{\bf x})\,d{\bf x},
 \end{equation}
 is singular. This implies that ${\bf e}^q=\nu_q{\bf f}^j$ for some $j$. 

One can trivially verify that the projection (\ref{eq_eig_proj}) has the
average  
 \begin{equation}
 s_q=\int s\,\rho^q(s)\,ds=\frac{1}{2}\delta_{q,0} 
 \end{equation}
 and the dispersion given by
 \begin{equation}
 \sigma_q^2=\int (s-s_q)^2\rho^q(s)\,ds
 =\frac{1}{24N}(1+\delta_{q,0}+\delta_{q,\pi})
 \end{equation}
 for all $q\ne q^*$ and almost always for $q=q^*$. A few degenerate cases
like $\rho({\bf x})=\delta(x_1-x_2)$ or $\rho({\bf x})=\delta(x_1+x_2-1)$
for $N=2$, or $\rho({\bf x})=\delta(x_1-x_3)\,\delta(x_2-x_4)$ for $N=4$
give different dispersions.

As expected, numerically calculating the projection $\rho^q(s)$ on the
stable and unstable directions (\ref{eq_evector}), we only get a singular
distribution for $q=q^*=0,\pi/3,\pi/2,2\pi/3,\pi$ (fig. 
\ref{fig_narrow}(b)), provided that the respective eigenvector is stable
($|\lambda_q|<1$). Otherwise a smooth Gaussian-like distribution is
obtained (fig. \ref{fig_narrow}(a)). 

Indeed one can easily see that in the large-length limit both the
continuous distribution and the envelope of the singular distribution
(\ref{eq_eig_proj}) become Gaussian:
 \begin{equation}
 \label{eq_gauss}
 \rho^q(s)\approx\cases{
    \frac{1}{\sigma_q}\phi\left(\frac{s-s_q}{\sigma_q}\right)
    & if $\forall j,\, {\bf e}^q\times{\bf f}^j\ne 0$, \cr
    \frac{\nu_q}{\sigma_q}\phi\left(\frac{s-s_q}{\sigma_q}\right)
    \delta(s-\nu_q p) & if $\exists j: {\bf e}^q=\nu_q{\bf f}^j$,
 }
 \end{equation}
 where $\phi(t)=(2\pi)^{-1/2}\exp(-t^2/2)$ is the normalized Gaussian and
$\nu_q=\kappa_q^{-1}$. 

\subsection{The Spatiotemporal Correlations}
\label{sec_spa_temp}

The standard spatial correlation function is trivially calculated to yield
 \begin{equation}
 \label{eq_corr_fun}
 C(r)=\langle x_i x_{i+r}\rangle-\langle x_i\rangle\langle x_{i+r}\rangle
 =\frac{1}{12}\,\delta_{r,0}
 \end{equation}
 for the completely expanding case with $\rho({\bf x})=1$ (here
$\langle\cdot\rangle$ denotes the average taken with $\rho({\bf x})$). 

If there are contracting directions, we rewrite (\ref{eq_corr_fun}) as
 \begin{eqnarray}
 C(r)&=&\sum_q(\sigma_{s,q}^2+\sigma_{c,q}^2)e^{iqr}\cr
 &=&\sum_q(\sigma_{s,q}^2+\sigma_{c,q}^2-\frac{1}{12N})e^{iqr}+
 \frac{1}{12}\,\delta_{r,0},
 \end{eqnarray}
 where $\sigma_{s,q}=\sigma_{c,q}=\sigma_q$ for all $q$ except
$\sigma_{s,0}=\sigma_{s,\pi}=0$. Since $\sigma_q^2=
(1+\delta_{q,0}+\delta_{q,\pi})/24N$ for all $q\ne q^*$,
 \begin{eqnarray}
 C(r)&=&\frac{1}{12}\,\delta_{r,0}+
 (\sigma_0^2-\frac{1}{12N})+(\sigma_\pi^2-\frac{1}{12N})(-1)^r\cr
 &+&2\sum_{q=\frac{\pi}{3},\frac{\pi}{2},\frac{2\pi}{3}}
 (\sigma_{s,q}^2+\sigma_{c,q}^2-\frac{1}{12N})\cos(qr).
 \end{eqnarray}

This reduces to a $\delta$-correlation (which coincides with the result
(\ref{eq_corr_fun}) obtained for $\rho({\bf x})=1$) in all but a few
special cases, when $\sigma_q^2 \ne (1+\delta_{q,0}+\delta_{q,\pi})/24N$.
For instance, choosing $m=0$ and $n=\pm 1$ yields for $N=4$
 \begin{equation}
 \rho({\bf x})=\delta(x_1-x_3)\delta(x_2-x_4)
 \end{equation}
 and therefore $\sigma_{s,\pi/2}=\sigma_{c,\pi/2}=0$ and
$\sigma^2_{c,0}=\sigma^2_{c,\pi}=1/24$, resulting in
 \begin{equation}
 \label{eq_corr_4}
 C(r)=\frac{1}{24}+\frac{(-1)^r}{24}=\cases{
    \frac{1}{12} & if $r=0,2$,\cr
    0 & if $r=1,3$.}
 \end{equation}

Since the invariant density, although being nontrivial, does not tell us
much about the spatiotemporal structures in the system, next we introduce
a spatiotemporal correlation function $G_i(r,t)$, which is a
straightforward generalization of the time correlation function
(\ref{eq_sing_cor}): 
 \begin{equation}
 \label{eq_time_cor}
 G_i(r,t) =\int d{\bf x}^0\rho({\bf x}^0)
 e^{2\pi i(x_i^0-x_{i+r}^t)}.
 \end{equation}

By expanding $\rho({\bf x})$ into Fourier series we obtain in analogy to
(\ref{eq_fou_comp}):
 \begin{equation}
 \label{eq_time_four}
 G_i(r,t)=\sum_{{\bf k}}\hat{\rho}({\bf k})
 \prod_{j=1}^N\delta(k_j-A_{i+r,j}^t+\delta_{i,j}).
 \end{equation}
 Since only the non-vanishing Fourier components $\hat{\rho}({\bf k}^*)=1$
(where ${\bf k}^*=\sum_l n_l{\bf f}^l$) of the invariant density
(\ref{eq_inv_dens}) contribute, (\ref{eq_time_four}) reduces to
 \begin{equation}
 G_i(r,t)=\sum_{n_1,\cdots,n_M}\prod_{j=1}^N\delta\left(
 \sum_{l=1}^M n_l f^l_j-A_{i+r,j}^t+\delta_{i,j}\right).
 \end{equation}
 In a translationally invariant system $G_i(r,t)$ does not depend
on $i$, so we drop the index and fix $i$ (set $i=1$ to be specific). 

It can be easily verified that the correlation (\ref{eq_time_cor}) is
short-ranged in both space and time. First we note that it vanishes if the
vector ${\bf k}_r^t$ with components $k_j=A_{1+r,j}^t-\delta_{1,j}$
does not lie on the stable manifold $W^s$. According to (\ref{eq_eig_exp})
 \begin{equation}
 A_{1+r,j}^t=\lambda_1^t\left(e^1_{1+r} e^1_j+\left(
 \frac{\lambda_2}{\lambda_1}\right)^t e^2_{1+r} e^2_j+\cdots\right),
 \end{equation}
 where $\lambda_1$ is the largest and $\lambda_2$ --- the next largest
eigenvalue and ${\bf e}^1$ and ${\bf e}^2$ are the respective
eigenvectors. For increasing $t$ the vector ${\bf k}_r^t$
asymptotically approaches the direction defined by ${\bf e}^1$ and
therefore cannot lie on the stable manifold for $t\ge\tau$, where $
\tau$ is some finite (and typically small) integer. 

On the other hand for $t=0$ we have
 \begin{equation}
 G(r,0)=\sum_{n_1,\cdots,n_M}\prod_{j=1}^N\delta\left(
 \sum_{l=1}^M n_l f^l_j-\delta_{1+r,j}+\delta_{1,j}\right).
 \end{equation}
 Since all basis vectors (\ref{eq_basis}) are periodic in space with
periods 1, 2, 3, 4 or 6, any linear combination of these will also be
periodic with period of at most 12. Since the vector with components $k_j=
\delta_{1+r,j}-\delta_{1,j}$ is {\em not} periodic for $r\ne 0$, the
maximal size of the system with non-trivial correlation is limited by
$N=12$. Again, choosing $m=0$ and $n=\pm 1$ for $N=4$ as an example, we
have ${\bf f}^1=(0,-1,0,1)$ and ${\bf f}^2=(1,0,-1,0)$ as basis
vectors and consequently
 \begin{equation}
 G(r,0)=\cases{1 & if $r=0,2$,\cr 0 & if $r=1,3$,}
 \end{equation}
 i.e. we retrieve the result (\ref{eq_corr_4}). 

\section{Discussion}
\label{sec_discuss}

To summarize, we have shown that the solution for the dynamical behavior
of a lattice of Bernoulli maps that are coupled by a matrix $A$ with
integer coefficients can be given in the closed form as ${\bf
x}^t=(A^t{\bf x}^0){\rm mod}\;1$:  the dynamical behavior of the coupled
map system can be described by the repeated action of a linear map
$A^t{\bf x}^0$ on variables that are confined to an $N$-torus. This
picture explains that the essentials of the dynamical behavior are
dictated by the eigenvalues and eigenvectors of $A$. 

The invariant density $\rho({\bf x})$ of this system displays Fourier
coefficients that are different from zero, i.e. is non-constant, whenever
the stable manifold of the zero wavevector contains a non-empty basis of
directions ${\bf f}^j$ with mutually rational components, generating the
infinite asymptotically contracting set of wavevectors. For nearest
neighbor couplings in a 1-dimensional lattice (given by eq. 
\ref{eq_mat_coup}) the maximal number of basis vectors is eight (actually
even less, since, e.g. $|\lambda_0-\lambda_\pi|=4|n|>2$).

We have calculated the standard spatial correlation function $C(r)$ for
the model with nearest neighbor couplings and shown that it is given by
$C(r)=\delta_{r,0}/12$ almost always. A few special cases exist however
for sufficiently small lattices, where the spatial correlations are
different. Nevertheless, $C(r)$ always vanishes at sufficiently large
distances. 

The invariant density of this system and the spatial correlation function
display little structure as compared to the Lyapunov spectrum, which is,
for the nearest neighbor coupling, given by $\Lambda_q=\log|m+2n\cos(q)|$.
This result shows that the time averaged spatial behavior is {\em not}
simply a straightforward reflection of the Lyapunov spectrum (see related
work listed in \cite{daido}).

We have also calculated the measurable spatiotemporal correlation function
$G(r,t)$ for the translationally invariant model and shown that it too is
short-ranged in both space and time.

It is instructive to compare our results with the general results obtained
by Bunimovich and Sinai \cite{sinai}, who proved that, for sufficiently
small coupling (in our case determined by parameter $\varepsilon$),
certain expanding coupled map systems with finite-range coupling possess
an absolutely continuous invariant measure $\mu({\bf x}):\ d\mu({\bf x})=
\rho({\bf x})\,d{\bf x}$, and also that the time and space correlation
functions decay exponentially (not slower than exponentially, to be
exact).

Our results indicate, that for larger coupling, the invariant measure
still exists, but might not be absolutely continuous due to the fact, that
large coupling often causes the appearance of contracting directions, even
if the isolated local maps $f(x)$ are expanding. The space and time
correlations in our model are seen to decay even faster then
exponentially, but the few special cases, giving non-trivial correlations
imply that there might be some general relationship between the continuity
of the invariant measure and the appearance of coherent structures in the
system.

Let us finally point out several directions of further research.

One open problem is the extension of our results to higher dimensions and
to couplings which have a longer range. In the 1-dimensional case the
eigenvectors remain also valid for longer-ranged couplings, only the
eigenvalues change. This means that a model with long, but finite, range
will have no more structure in the invariant density than the short-ranged
model. This is of course a peculiarity of the Bernoulli shift map, but
should again be taken as a warning for making conclusions from the spatial
range of the coupling onto the observable spatial patterns. 

Although our solution for the dynamics and the correlation functions hold
for general dimensions it would be interesting to see what the
restrictions on the wavevectors that generate the basis of the invariant
density look like in two and three dimensions.

Finally, one could investigate the dynamical behavior of a system, whose
time dependence is given a priori by the equation (\ref{eq_solution}) 
also for matrices $A$ with {\em non-integer} elements. By doing so one
will loose the property of the original map, that the relation relation
(\ref{eq_map_vec}) holds step by step, but the trajectories generated by
equation (\ref{eq_solution}) are well defined. 

\acknowledgements

H.G.S. thanks C. Koch for the kind hospitality extended to him at Caltech
and the Volkswagen Foundation for financial support. The authors thank M. 
C.  Cross for the careful reading of the manuscript. This research has
also been partially supported by the NSF through grant No. DMR-9013984.

\end{document}